\documentclass[11pt]{article}
\usepackage{graphicx}
\usepackage[T1]{fontenc} 
\usepackage[english]{babel} 

\usepackage[left=2.5cm,top=4.5cm,right=2.5cm,bottom=3cm]{geometry}

\usepackage{mathptmx}
\usepackage[scaled=.85]{helvet}

\usepackage{amsfonts}
\usepackage{amssymb}
\usepackage{amsmath}
\usepackage{amsthm}
\usepackage[justification=centering]{caption}

\usepackage{abstract}

\usepackage{titling}
\posttitle{\par\end{center}}

\usepackage{titlesec}
\titleformat*{\section}{\fontsize{16pt}{18pt}\selectfont \bfseries}
\titleformat*{\subsection}{\fontsize{14pt}{8pt}\selectfont \bfseries}
\titleformat*{\subsubsection}{\fontsize{12pt}{8pt}\selectfont \bfseries}

\theoremstyle{plain}

\theoremstyle{remark}

\usepackage{enumitem}
\setlist[itemize]{noitemsep}

\usepackage{etoolbox}
\apptocmd{\thebibliography}{\setlength{\itemsep}{-2pt}}{}{}

\usepackage[hidelinks]{hyperref}
\hypersetup{
  pdftitle={Evaluating Soccer Player Movements Using the Attacker-Defender Model},
  pdfauthor={Takuma Narizuka and Issei Yamazaki}
}

\usepackage{setspace}

\usepackage{fancyhdr}
\newcommand{\pvec}[1]{\vec{#1}\mkern2mu\vphantom{#1}}
\usepackage{float}
\usepackage{placeins}

\usepackage[none]{hyphenat}

\title{\fontsize{20pt}{22pt}\selectfont
{\bf Evaluating Soccer Player Movements Using the Attacker-Defender Model}}

\author{\vspace{8pt}
Takuma Narizuka* and Issei Yamazaki**\\
\fontsize{10pt}{6pt}\selectfont
*Faculty of Data Science, Rissho University, Kumagaya, Saitama, Japan, narizuka@ris.ac.jp\\
\fontsize{10pt}{6pt}\selectfont
**Meiji Institute for Advanced Study of Mathematical Sciences, Meiji University, Nakano-ku, Tokyo, Japan
}

\date{}

\begin{document}

\pagenumbering{gobble}

\maketitle

\vspace{-30pt}

\begin{abstract}
The present study investigates the attacker-defender (AD) model proposed by Brink et al. (2023), a motion model that describes the interactions between a ball carrier (attacker) and the nearest defender during ball possession.
The model is based on the equations of motion for both players, incorporating resistance, goal-oriented force, and opponent-oriented force.
It generates trajectories based on physically interpretable parameters.
Although the AD model reproduces real dribbling trajectories well, previous studies have explored only a limited range of parameter values and relied on relatively small datasets.
This study aims to (1) enhance parameter optimization by solving the AD model for one player with the opponent's actual trajectory fixed, (2) validate the model's applicability to a large dataset from 306 J1 League matches, and (3) demonstrate distinct playing styles of attackers and defenders based on the full range of optimized parameters.
\end{abstract}

\section{Introduction}
With the widespread availability of tracking and event data \cite{Bassek2025}, the generation of short-term player trajectories has received increasing attention in football (soccer) analytics.
Machine learning-based approaches \cite{Brefeld2019} focus on predictive accuracy, while physics-based models \cite{Fujimura2005} prioritize interpretability.
The present study adopts a physics-based motion model, incorporating fundamental principles of player behavior.
Among one-dimensional motion models, the Keller model \cite{Keller1973} has been widely used to analyze sprinting, where velocity evolves as $ dv/dt = -v(t)/\tau + f$.
However, the analysis of soccer player movements necessitates the use of two-dimensional models.
A notable example is the Fujimura-Sugihara model \cite{Fujimura2005}, an extension of the Keller model, which has been validated through tracking data \cite{Narizuka2023}.
In a recent study, Brink et al. \cite{Brink2023} proposed the Attacker-Defender (AD) model, which incorporates the interaction between a ball carrier (attacker) and the nearest defender into the Fujimura-Sugihara model.
The model generates trajectories based on physically interpretable parameters and reproduces a wide range of dribbling movements.
However, previous studies have primarily explored a limited range of parameter values using small datasets, and the model's applicability to large datasets remains underexplored.
The present study has three primary objectives.
First, we enhance the parameter optimization process of the AD model.
Specifically, we propose a method that solves the model for one player with the opponent's actual trajectory fixed.
Second, we validate the model's applicability using a significantly larger dataset than those employed in previous studies.
We analyze tracking and event data from 306 J1 League matches provided by DataStadium Inc., Japan.
Finally, we quantitatively identify characteristic ball carriers (attackers) and their nearest defenders based on the optimized parameters of the AD model.
By expanding the range of analyzed parameters, we provide new insights into the playing styles of attackers and defenders.

\section{Methods}

\subsection{Dataset}
We used data from 306 matches in the Japan Professional Football League (J1 League) during the 2023 season, which was provided by DataStadium Inc., Japan.
The dataset contains the absolute position coordinates $ (x, y) $ of all players, recorded every 0.04 seconds.
The dataset also includes event data, such as timestamps for ball possession, passes, and shots.
The Savitzky--Golay filter and cubic spline interpolation were applied to the position data to reduce noise, and velocities were subsequently computed from the smoothed trajectories (see \cite{Brink2023} for details).

For the analysis of dribbling situations, all instances of dribbling were extracted and the corresponding tracking data for the ball carrier and the nearest defender were identified.
We further selected only the dribbling events that satisfied the following criteria: (i) the nearest defender did not change during the dribble, (ii) the duration of ball possession exceeded 0.5 seconds, and (iii) the linear distance traveled by both players exceeded 5 meters.
Consequently, 31,028 instances of dribbling were obtained for parameter optimization, which is significantly larger than the 1,573 dribbling events analyzed in the previous study \cite{Brink2023}.
\begin{figure}[htbp]
  \centering
  \includegraphics[width=1\linewidth]{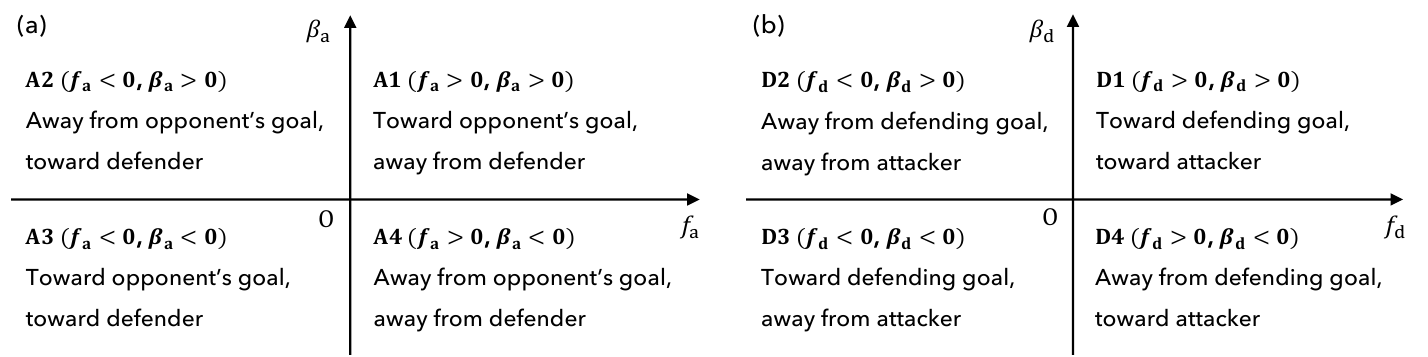}
  \caption{Expected attacker and defender motions based on the signs of (a) $f_{\mathrm{a}}$ and $\beta_{\mathrm{a}}$, and (b) $f_{\mathrm{d}}$ and $\beta_{\mathrm{d}}$.
  }
  \label{fig:motion_classification}
\end{figure}

\subsection{Attacker-Defender Model}
The Attacker-Defender (AD) model describes the movement of an attacker and a defender during a single dribbling event.
Let $ \vec{v}_{\mathrm{a}} $ and $ \vec{v}_{\mathrm{d}} $ denote the velocity vectors of the attacker and defender, respectively.
The equations of motion are as follows \cite{Brink2023}:
\begin{align}
  \frac{d\vec{v}_{\mathrm{a}}}{dt} &= -\frac{1}{\tau_{\mathrm{a}}} \vec{v}_{\mathrm{a}}(t) + f_{\mathrm{a}} \frac{\beta_{\mathrm{a}} \vec{e}_{\mathrm{ag}}(t) - \vec{e}_{\mathrm{ad}}(t)}{\left\lVert \beta_{\mathrm{a}} \vec{e}_{\mathrm{ag}}(t) - \vec{e}_{\mathrm{ad}}(t) \right\rVert}, \label{AT_model_a} \\[8pt]
  \frac{d\vec{v}_{\mathrm{d}}}{dt} &= -\frac{1}{\tau_{\mathrm{d}}} \vec{v}_{\mathrm{d}}(t) + f_{\mathrm{d}} \frac{\beta_{\mathrm{d}} \vec{e}_{\mathrm{dg}}(t) + \vec{e}_{\mathrm{da}}(t)}{\left\lVert \beta_{\mathrm{d}} \vec{e}_{\mathrm{dg}}(t) + \vec{e}_{\mathrm{da}}(t) \right\rVert}. \label{AT_model_d}
\end{align}
Here, the first term on the right-hand side represents the resistance force proportional to velocity.
The second term represents a driving force determined by the goal-oriented component and the opponent-related component.
For positive $f$ and $\beta$, this term drives the attacker toward the opponent's goal and away from the defender, whereas it drives the defender toward the defending goal and toward the attacker.
Thus, the total force exerted by the attacker and defender is determined by the balance of the resistance force and these driving components.
In the model parameters, $ f_{\mathrm{a}} $ and $ f_{\mathrm{d}} $ represent the maximum driving forces of the attacker and defender, respectively, while $\tau_{\mathrm{a}} $ and $\tau_{\mathrm{d}} $ denote the time constants required to reach their maximum speed.
The parameters $\beta_{\mathrm{a}}$ and $\beta_{\mathrm{d}}$ control the relative weights of the goal-oriented and opponent-oriented driving forces.
The trajectories of the attacker and defender were obtained by numerically solving the coupled equations of motion with appropriate initial conditions.
Note that these trajectories depend on the six parameters in the model.
Figures~\ref{fig:motion_classification}(a) and~(b) illustrate the classification of expected attacker and defender motions based on the signs of $f_{\mathrm{a}}$ and $\beta_{\mathrm{a}}$, and $f_{\mathrm{d}}$ and $\beta_{\mathrm{d}}$, respectively.
While the previous study \cite{Brink2023} limited its analysis to regions A1 and D1, the present study expands it to all four quadrants.

\subsection{Parameter Optimization}
Let $ \vec{r}_{\mathrm{p}}(t) $ and $ \pvec{r}'_{\mathrm{p}}(t) $ denote the actual and simulated trajectories of player $ \mathrm{p} \in \{\mathrm{a}, \mathrm{d}\} $, respectively.
Following \cite{Brink2023}, the trajectory error is defined as
\begin{align}
\varepsilon_{\mathrm{p}} = \frac{1}{T}\frac{\sum_{t=0}^{T-1} \|\vec{r}_{\mathrm{p}}(t) - \pvec{r}'_{\mathrm{p}}(t)\|}{\sum_{t=1}^{T-1} \|\vec{r}_{\mathrm{p}}(t) - \vec{r}_{\mathrm{p}}(t-1)\|},
\label{eq:error}
\end{align}
where $ T $ is the number of frames in a single dribbling event.
The error values $\varepsilon_{\mathrm{a}}$ and $\varepsilon_{\mathrm{d}}$ are functions of the model parameters.
In the original coupled approach, the six model parameters are optimized by minimizing the total error defined as $\varepsilon = \varepsilon_{\mathrm{a}} + \varepsilon_{\mathrm{d}}$.
In the improved approach used for the player characterization in this study, the parameters of each player were optimized independently: $(f_{\mathrm{a}}, \tau_{\mathrm{a}}, \beta_{\mathrm{a}})$ were estimated by minimizing $\varepsilon_{\mathrm{a}}$ with the defender's actual trajectory fixed, and $(f_{\mathrm{d}}, \tau_{\mathrm{d}}, \beta_{\mathrm{d}})$ were estimated by minimizing $\varepsilon_{\mathrm{d}}$ with the attacker's actual trajectory fixed.
The combined error $\varepsilon = \varepsilon_{\mathrm{a}} + \varepsilon_{\mathrm{d}}$ was then used to evaluate the overall reproduction accuracy of each dribbling event.
Parameter optimization was performed for each dribbling event using the COBYLA algorithm from the SciPy library in Python.
For each event, initial parameter values were generated by Latin hypercube sampling with
$ f_{\mathrm{a}}, f_{\mathrm{d}} \in [-11.3, 11.3] $,
$ \beta_{\mathrm{a}}, \beta_{\mathrm{d}} \in [-5, 5] $, and
$ \tau_{\mathrm{a}}, \tau_{\mathrm{d}} \in [0.9, 2.5] $.
During optimization, the parameters were constrained as
$ \tau_{\mathrm{a}}, \tau_{\mathrm{d}} \ge 0.9 $,
$ \beta_{\mathrm{a}}, \beta_{\mathrm{d}} \in [-10, 10] $, and
$ |f_{\mathrm{a}}\tau_{\mathrm{a}}|, |f_{\mathrm{d}}\tau_{\mathrm{d}}| \le 10.2 $.
For each event, $N$ random initial parameter sets were generated, and the one yielding the lowest relevant error was selected.

\section{Results}

\subsection{Improving Parameter Optimization}
In the previous study \cite{Brink2023}, the equations of motion \eqref{AT_model_a} and \eqref{AT_model_d} were solved simultaneously to generate the trajectories of both players.
We propose a new method that solves the equations of motion for the attacker and defender independently, using the actual trajectory of the other player.
This enables the independent parameter optimization for attacker and defender.
Consequently, the dimensions of parameter space are reduced, thereby significantly decreasing the computational cost.
To estimate optimized parameters of 31,028 dribbling events, Latin hypercube sampling was employed to generate $ N = 100 $ initial parameter sets.
In the original approach, the set minimizing $\varepsilon$ was selected for each event, whereas in the improved approach, the sets minimizing $\varepsilon_{\mathrm{a}}$ and $\varepsilon_{\mathrm{d}}$ were selected independently.
We then retained events where both  $ \varepsilon_{\mathrm{a}} $ and  $ \varepsilon_{\mathrm{d}} $ were below 0.1.
Consequently, 26,492 events (85.4\%) met this criterion using the original method, whereas 27,457 events (88.5\%) did so using the improved approach.
The proposed approach thus not only reduced the computational cost, but also increased the number of successful parameter estimates.

\subsection{Attacker Characterization}
\begin{figure*}[htbp]
    \centering
    \includegraphics[width=\linewidth]{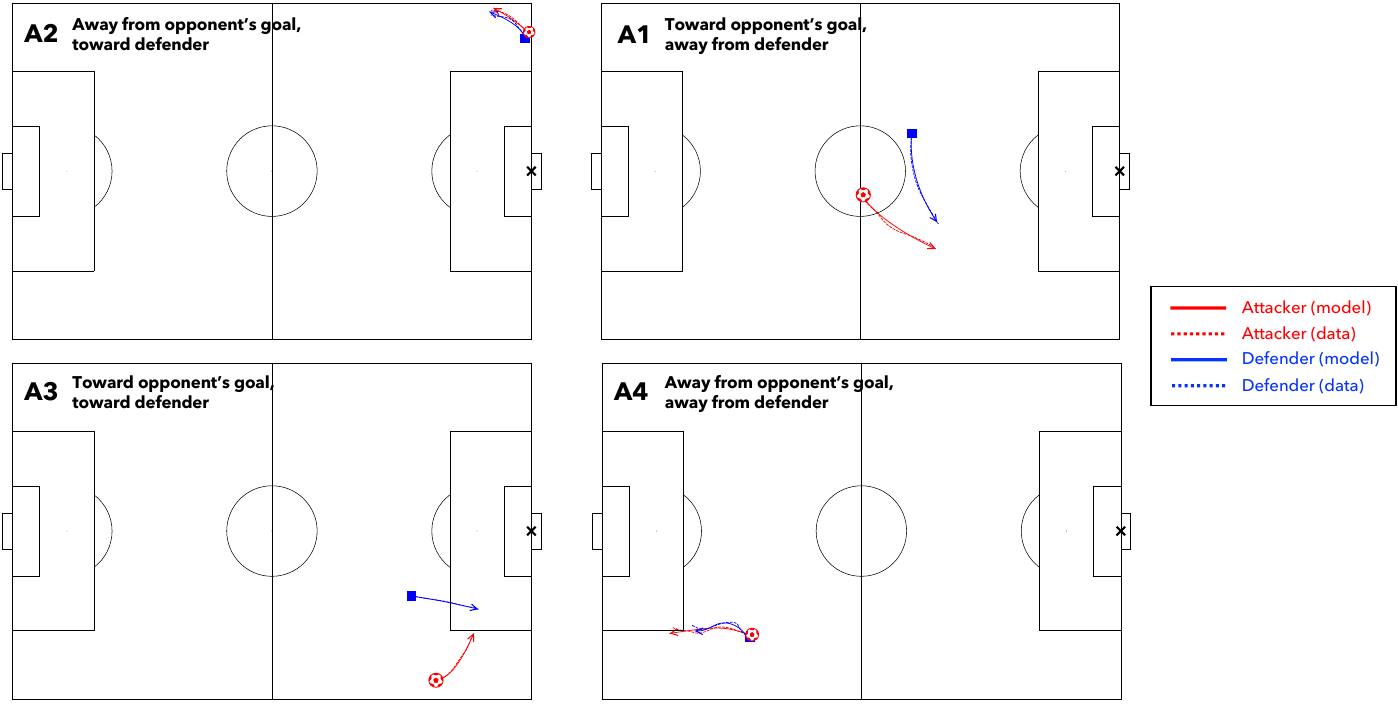}
    \caption{Typical attacker trajectories in regions A1--A4. The red circle and the blue square indicate the starting positions of the attacker and defender, respectively. Dashed and solid lines represent the actual and simulated trajectories, respectively. In all cases, the direction of attack is from left to right.}
    \label{fig:attacker_trj}
\end{figure*}

The improved approach reproduced 27,457 events with both errors below 0.1.
The player characterization was therefore conducted using these 27,457 accurately reproduced trajectories.
The attacker trajectories were categorized into four quadrants based on the signs of the optimized parameters, $ f_{\mathrm{a}} $ and $ \beta_{\mathrm{a}} $.
Table~\ref{tab:attacker_example} lists the most frequently appearing attackers in each quadrant.
In region A1 ($f_{\mathrm{a}} > 0,\ \beta_{\mathrm{a}} > 0$), the AD model suggests that the attacker tends to move toward the opponent's goal and away from the defender.
Two typical patterns were observed: maintaining a safe distance while advancing the ball, and drawing the defender in before releasing the ball (Fig.~\ref{fig:attacker_trj}, A1).
Center backs engaged in building-up play were frequently observed in this region.
In region A2 ($f_{\mathrm{a}} < 0,\ \beta_{\mathrm{a}} > 0$), the attacker is modeled to move away from the opponent's goal and toward the defender.
The observed situations included maintaining possession under pressure from behind, and dribbling directly toward a nearby defender (Fig.~\ref{fig:attacker_trj}, A2).
In region A3 ($f_{\mathrm{a}} < 0,\ \beta_{\mathrm{a}} < 0$), the parameter values imply movement toward both the opponent's goal and the defender.
Typical cases involved direct one-on-one confrontations or forward progression toward the defender (Fig.~\ref{fig:attacker_trj}, A3).
The dribbling in this region corresponds to actions near scoring chances, which are crucial for evaluating attackers.
Finally, in region A4 ($f_{\mathrm{a}} > 0,\ \beta_{\mathrm{a}} < 0$), the model generates movement away from both the opponent's goal and the defender.
The observed patterns included forward acceleration under rear pressure and backward dribbles that drew in the defender (Fig.~\ref{fig:attacker_trj}, A4).
\begin{table*}[htbp]
    \centering
    \caption{Examples of the most frequently appearing attackers in regions A1--A4. The values in parentheses indicate the player's position and the number of occurrences.}
    \label{tab:attacker_example}
    \scalebox{0.95}{
    \begin{tabular}{llll}
    \hline
    A1 & A2 & A3 & A4 \\ \hline
    A. Scholz (DF, 290) & A. Ienaga (FW, 49) & J. Alano (MF, 29) & J. Schmidt (MF, 27)  \\
    M. H\o ibr\aa ten (DF, 258) & M. H\o ibr\aa ten (DF, 38) & \'Elber (FW, 27) & K. Watanabe (MF, 24)  \\
    T. Deng (DF, 248) & A. Scholz (DF, 37) & T. Okubo (MF, 22) &  S. Kagawa (MF, 17) \\ 
    Eduardo (DF, 200) & K. Yuruki (MF, 32) & Y. Muto (FW, 22) & Y. Wakizaka (MF, 17)\\ 
    D. Okamura (DF, 194) & Y. Matheus (FW, 31) & K. Anzai (DF, 22)  & S. Kawahara (MF, 16) \\ \hline
    \end{tabular}
    }
\end{table*}

\FloatBarrier

\subsection{Defender Characterization}
\begin{figure*}[htbp]
    \centering
    \includegraphics[width=\linewidth]{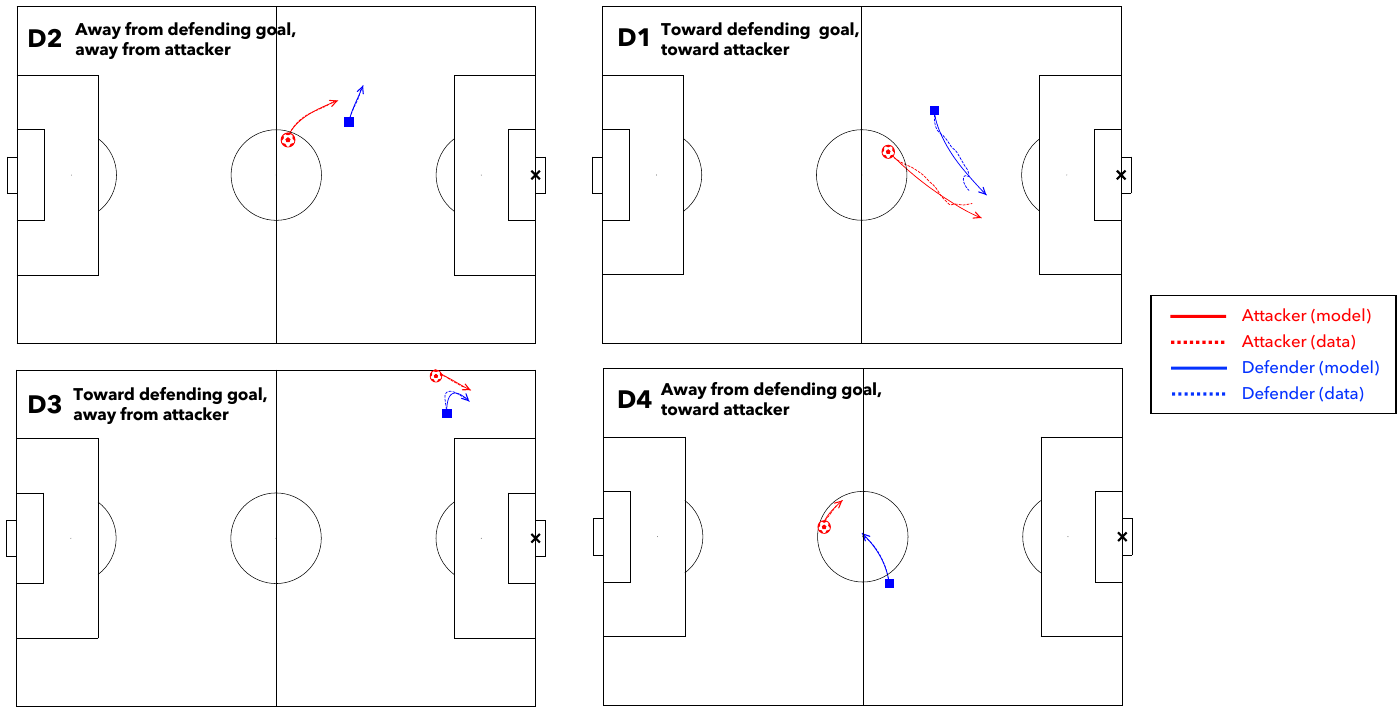}
    \caption{Typical defender trajectories in regions D1--D4. The red circle and the blue square indicate the starting positions of the attacker and defender, respectively. Dashed and solid lines represent the actual and simulated trajectories, respectively. In all cases, the direction of attack is from left to right.}
    \label{fig:defender_trj}
\end{figure*}

We also characterized defender behavior across the four quadrants using the 27,457 accurately reproduced defender trajectories.
Table~\ref{tab:defender_example} lists the most frequently appearing players in each quadrant, and Fig.~\ref{fig:defender_trj} shows typical movement patterns.
In region D1 ($f_{\mathrm{d}} > 0,\ \beta_{\mathrm{d}} > 0$), the model generates movement toward both the defending goal and the attacker.
We observed a pressing action in which the defender retreated toward his goal while guiding the attacker toward the sideline.
Center forwards of defending teams were frequently observed in this region.
In region D2 ($f_{\mathrm{d}} < 0,\ \beta_{\mathrm{d}} > 0$), the model generates movement away from both the defending goal and the attacker.
This region often included defensive retreat to prevent direct dribbling toward the goal.
In region D3 ($f_{\mathrm{d}} < 0,\ \beta_{\mathrm{d}} < 0$), the model generates movement toward the defending goal and away from the attacker.
Typical defensive behavior included dealing with dribbles coming in from the side.
Finally, in region D4 ($f_{\mathrm{d}} > 0,\ \beta_{\mathrm{d}} < 0$), the model generates movement away from the defending goal and toward the attacker.
This region included cases where defenders challenged attackers from behind, typically under high-risk situations.
Because actions in this region can lead to ball recovery or changes of possession, it is crucial for evaluating defenders.
\begin{table*}[htbp]
    \centering
    \caption{Examples of the most frequently appearing defenders in regions D1--D4. The values in parentheses indicate the player's position and the number of occurrences.}
    \label{tab:defender_example}
    \scalebox{0.95}{
    \begin{tabular}{llll}
    \hline
    D1 & D2 & D3 & D4 \\ \hline
    M. Hosoya (FW, 354) & M. Hosoya (FW, 46) & M. Hosoya (FW, 29) & M. Hosoya (FW, 34)  \\
    L. Cear\'a (FW, 264) & Y. Yamagishi (FW, 22) & Y. Muto (FW, 16) & Y. Osako (FW, 21)  \\
    K. Junker (FW, 262) & Y. Matheus (FW, 21) & I. Jebali (FW, 15) & Y. Suzuki (FW, 20) \\ 
    Y. Yamagishi (FW, 247) & T. Kikuchi (MF, 21) & L. Cear\'a (FW, 14) & T. Miyashiro (FW, 19) \\ 
    Y. Suzuki (FW, 229) & Y. Muto (FW, 21) & S. Kagawa (MF, 13) & Y. Yamagishi (FW, 18) \\ \hline
    \end{tabular}
    }
\end{table*}

\section{Conclusions and Future Work}
The present study enhanced the parameter optimization of the Attacker-Defender model and validated its applicability using a large dataset.
The equations of motion for the attacker and defender were solved independently, using the actual trajectory of the other player.
This modification reduced computation cost and increased the number of dribbles accurately reproduced by the model.
In addition, using the 27,457 accurately reproduced dribbling events, player trajectories were categorized into four quadrants based on the optimized parameters.
Each quadrant exhibited distinct playing styles, and the most frequently appearing players were identified.
Future studies will focus on refining the error function and conducting more detailed analyses of dribbling events.
The optimized parameters often approached the boundaries of the parameter space, indicating a need for further refinement of the error function.
For example, incorporating penalty terms based on player velocity and acceleration could be considered.
Additionally, analyzing dribbling events in specific contexts (e.g., actions on each side of the field) or for specific player roles (e.g., midfielders or forwards) may yield more detailed insights.

\section*{Acknowledgements}
The authors are grateful to DataStadium Inc., Japan, for providing the data used in this study.
This research was supported in part by the Data-Centric Science Research Commons Project of the Research Organization of Information and Systems, Japan, a Grant-in-Aid for Early-Career Scientists (No.23K16729) from the Japan Society for the Promotion of Science (JSPS).

\end{document}